\documentclass[a4paper,pre,superscriptaddress,floatfix,nofootinbib,twocolumn]{revtex4-1}

\usepackage{bbold}
\usepackage{bbm}
\usepackage[pdftex]{graphicx}
\usepackage{latexsym,amsmath,verbatim,amssymb,txfonts}
\usepackage{color}
\usepackage{rotating}
\usepackage{verbatim}
\usepackage{multirow}
\usepackage[english]{babel}
\usepackage{comment}
\usepackage{hyperref}

\newcommand{\change}[1]{{#1}}

\begin{document}

\title{Combinatorial approach to spreading processes on networks}

\author{Dario Mazzilli}
\affiliation{Center for Complex Networks and Systems Research, Luddy School
  of Informatics, Computing, and Engineering, Indiana University, Bloomington,
  Indiana 47408, USA}%
\author{Filippo Radicchi}%
 \email{filiradi@indiana.edu}
\affiliation{Center for Complex Networks and Systems Research, Luddy School
  of Informatics, Computing, and Engineering, Indiana University, Bloomington,
  Indiana 47408, USA}%


\begin{abstract}
Stochastic spreading models defined on complex network topologies are used to mimic the diffusion of diseases, information, and opinions in real-world systems.
Existing theoretical approaches 
to the characterization of the models in terms of 
microscopic configurations
rely on some approximation of independence among dynamical variables, thus introducing a systematic bias in the prediction of the ground-truth dynamics. 
Here, we develop a combinatorial framework based on the approximation that spreading may occur only along the shortest paths connecting pairs of nodes.
The approximation overestimates dynamical correlations among node states and leads to biased predictions. Systematic bias is, however, pointing in the opposite direction of existing approximations.  We show that the combination of the two biased approaches generates predictions of the ground-truth dynamics that are more accurate than the ones given by the two approximations if used in isolation. We further take advantage of the combinatorial approximation to characterize theoretical properties of some inference problems, and show that the reconstruction of microscopic configurations is very sensitive to both the place where and the time when partial knowledge of the system is acquired.
\end{abstract}

\maketitle


\section{\label{sec:Intro}Introduction}

Stochastic spreading models running on top of network topologies have been used to study a large variety of real-world dynamical processes~\cite{pastor2015epidemic,butts2009revisiting,jackson2010social,vespignani2012modelling}. Examples include the spread of diseases \cite{lloyd2001viruses,eames2002modeling}, the diffusion of information and opinions \cite{weng2014predicting,castellano2009statistical,moreno2004dynamics,dall2006nonequilibrium}, the propagation of bank failures \cite{brandi2018epidemics}, blackout cascades \cite{dobson2016obtaining}, development of countries \cite{hidalgo2007product}, and \change{avalanche dynamics in} neural networks \cite{vogels2005neural}.

In spite of their simplicity, many spreading models can be solved exactly on very specific network topologies only; extensive simulations and/or theoretical approximations  are  generally  required  to  characterize  their  properties on  arbitrary  networks~\cite{pastor2015epidemic}. A complete solution of a spreading model on a network consists in associating a probability to every possible microscopic configuration of the system at each instant of time. Such a detailed knowledge may not be required in  applications where the interest is centered around the macroscopic behavior of the system, e.g., outbreak size and/or duration~\cite{moreno2002epidemic,payne2011exact,castellano2010thresholds,buzna2006modelling}. It is however required in many other applications of central importance for spreading processes on networks as for example the problems of inferring the patient zero identity ~\cite{altarelli2014bayesian, lokhov2014inferring}, optimal sampling~\cite{radicchi2018uncertainty}, and influence maximization \cite{kempe2003maximizing}. 
Existing theoretical approaches to the microscopic description of spreading processes on networks rely on some approximation of independence among the state variables of the individual nodes. For example, the individual-node mean-field approximation assumes complete dynamical independence among state variables of the individual nodes~\cite{wang2003epidemic, Chakrabarti_2008, pastor2015epidemic}. Message-passing approximations, such as those considered in Refs.\change{~\cite{karrer2010message, lokhov2015dynamic,cator2014nodal,gleeson2013binary}}, improve over the mean-field approximation by relying on conditional independence among pairs of variables, providing exact predictions on tree-structured networks and excellent predictive power on arbitrary networks. These approximations have the common bias of neglecting, to some extent, dynamical correlations that are essential \change{for the exact description of} unidirectional spreading processes. As a consequence, their predictions are systematically biased towards the overestimation of the infection probability of individual nodes. From the computational point of view, these approximations allow for the quick computation of marginal probabilities. However, they lack of flexibility with respect to changes in the dynamical and/or topological details of the process. If the initial conditions of the dynamics, the parameter values of the spreading model, or the topology of the network are changed, solutions of the approximations should be computed afresh by iteration.

In this paper, we introduce an approximation based on a combinatorial calculation of the spreading probability along the shortest path between pairs of nodes. The approximation neglects that an infection may propagate along paths longer than the shortest one. We derive close-form expressions of the approximation for the Susceptible-Infected and Susceptible-Infected-Recovered models started from a single source of infection. On trees, our approach is exact, providing a geometric interpretation of correlations and joint probabilities between pairs of nodes. We leverage such an intuitive interpretation of the approach to study properties of the patient-zero problem and to compare  different strategies of acquiring information about network configurations from partial observations. 
In networks with loops, the approach systematically underestimates the probability of infection of individual nodes. The bias goes in the opposite direction of existing approximations for spreading processes on networks. The simultaneous use of the two types of approximations allows us to define a region where the true value of probabilities lies. Their combination improves the accuracy of each individual approximation. We stress that the computationally demanding component of our approximation consists in finding the shortest path between  pairs of nodes. Probabilities of node states are then determined solely on such a geometric knowledge.  As a result, exploring the parameters' space of a spreading model is very efficient. We are aware of existing approaches that approximate spreading as happening on the shortest paths among pairs of nodes only~\cite{brockmann2013hidden}. However, we are not aware of a full theoretical development of the approximation, consisting of closed-form combinatorial solutions, as the one presented here.

The paper is organized as follows. In section~\ref{sec:models}, we introduce the spreading models considered in the paper. We further describe the individual-based mean-field approximation and our combinatorial approximation. We compare the accuracy of the approximations in predicting ground-truth spreading dynamics. In the comparisons, we include also the dynamic message-passing approximation. In section~\ref{sec:applications}, we apply our approximation to trees, and characterize some properties of inference problems associated to spreading, including the identification of the patient zero and the maximization of system information from the local observation of the states of some nodes. In section~\ref{sec:conclusions}, we summarize our results and indicate viable extensions of our work.

\section{\label{sec:models}Theoretical approximations for spreading dynamics on networks}


We assume that spreading occurs on a quenched, undirected and unweighted, network composed of $N$ nodes. The topology of the network is
fully specified by the $N \times N$ adjacency matrix $A$, whose
generic element $A_{ij} = A_{ji} = 1$ if nodes $i$ and $j$ are connected, whereas $A_{ij}= A_{ji} = 0$, otherwise. We assume that the network does not contain any self-connection, so that $A_{ii} = 0$ , $\forall i$.
Without loss of generality, we further assume that the network is composed of a single connected component, so that every node is reachable from any other node, and spreading from any single initial source node has the potential to involve the entire system.
We indicate with $\ell_{ij}$ the distance between nodes $i$ and $j$ in the network, equal to the minimal number of edges that separate the two nodes. Please note that the symmetry of the network implies that $\ell_{ij} = \ell_{ji}$.

We consider the discrete-time version of two very popular models of spreading dynamics: the  Susceptible-Infected (SI) and the Susceptible-Infected-Recovered (SIR) models \cite{newman2002spread,anderson1992infectious,pastor2015epidemic}. 

In the SI model, every node $i$ at time $t$ can be found in two different states, either the susceptible state $\sigma_i^{(t)} = S$ or the infected state $\sigma_i^{(t)} = I$. At each discrete stage of the dynamics $t>0$, 
every node $i$ such that $\sigma_i^{(t-1)} = I$
tries to infect every neighbor $j$, i.e., $A_{ij}=1$, in the susceptible state. The infection is successfully transmitted with spreading probability $0 \leq \beta \leq 1$. A successful spreading event consists in changing the state of node $j$ 
as $\sigma_j^{(t-1)} = S \to \sigma_j^{(t)} = I$. 
This means that the newly infected node $j$ can attempt to further spread the infection from time $t+1$ on. After all spreading attempts for all infected nodes have been considered, time increases as $t \to t +1$. The dynamics is such that, as long as at least one node is initially infected and $\beta > 0$, all nodes will, sooner or later, end up in the infected state. 

The SIR model is a slightly more sophisticated model than the SI model, and the generic node $i$ may be also found in the recovered state $\sigma_i^{(t)} = R$. Spreading events happen in the same exact way as for the SI model. However, after all spreading attempts of stage $t$ have been considered, then every node $i$ such that $\sigma_i^{(t-1)} = I$ may recover with probability $0 \leq \gamma \leq 1$. Recovery of node $i$ consists in the change of state 
$\sigma_i^{(t-1)} = I \to \sigma_i^{(t)} = R$. 
After all recovery attempts have been considered, time increases as $t \to t +1$. Recovered nodes do not participate in the spreading dynamics, in the sense that they cannot infect their susceptible neighbors nor they can be re-infected by their infected neighbors. Potentially many different final configurations are reachable depending on the choice of the parameters $\beta$ and $\gamma$, and the initial configuration of the system.


\subsection{\label{sec:SI}The susceptible-infected model}

In this section, we derive  analytical expressions for the SI model on arbitrary network topologies. We will first consider the individual-based mean-field approximation (IBMFA)~\cite{wang2003proceedings,Chakrabarti_2008, pastor2015epidemic}. Then, we will derive a novel approximation based on combinatorial arguments. The novel approximation is exact on trees, and is expected to perform well on sparse tree-like networks. We name the method as the shortest-path combinatorial approximation (SPCA). We will characterize some properties of SPCA, and compare the prediction accuracy of the novel approximation against IBMFA and the so-called dynamic message-passing approximation (DMPA)~\cite{lokhov2015dynamic}. DMPA is the best approximation on the market for the prediction of marginal probabilities in the SI (and SIR) model. However, given their complicated form, we will not report DMPA equations below. The interested reader can find the equations, and their derivation, in Ref.~\cite{lokhov2015dynamic}.

\subsubsection*{Problem setting}
We assume that spreading is initiated by a single infected node $s$, i.e., $\sigma_{s}^{(0)} = I$. Node $s$ is the source of the infection or the patient zero. All other nodes $j$ are initially in the susceptible state, i.e., $\sigma_{\forall j \neq s}^{(0)} = S$. Our goal is to fully characterize the probability of the microscopic state of  every individual node at each stage of the dynamics. The main quantity that we focus on is
\begin{equation}
    P_{s\to i}^{(t)} = \textrm{Prob.} \left[ \sigma_{i}^{(t-1)} = S \to \sigma_{i}^{(t)}=I \left| \sigma_{\forall j \neq s}^{(0)} =S, \sigma_{s}^{(0)}=I \right. \right] \; ,
    \label{eq:P}
\end{equation}
i.e., the probability that the infection, started from the source node $s$ at time $t=0$, reaches node $i$ after exactly $t$ stages of the dynamics. We will consider different expressions for  $P_{s\to i}^{(t)}$ on the basis of the above-mentioned approximations. Once 
$P_{s\to i}^{(t)}$ is given, several other quantities useful in the characterization of the model dynamics can be immediately computed. For example, to obtain the probability $Q_{s \to i}^{(t)}$ that node $i$ has been infected at time $t$ or earlier, we simply perform the sum
\begin{equation}
    Q_{s \to i}^{(t)} = \sum_{r=0}^{t} P_{s \to i}^{(r)} \; .
    \label{eq:Q_sing}
\end{equation}

 Based on our assumption on the initial configuration, we automatically have that $P_{s \to s}^{(0)} = Q_{s \to s}^{(0)} = 1$, and $P_{s \to i}^{(0)} = Q_{s \to i}^{(0)} = 0$  $\; , \forall \, i \neq s$.

$P_{s\to i}^{(t)}$ and $Q_{s \to i}^{(t)}$ are  probabilities subjected  to the initial condition that spreading is initiated by node $s$. We can relax such a condition and consider arbitrary initial configurations consisting of one unknown source. If we indicate with 
\begin{equation}
z_s = \textrm{Prob.} \left[ \sigma_{\forall j \neq s}^{(0)} =S, \sigma_{s}^{(0)}=I  \right]
\label{eq:prior}
\end{equation}
the probability that node $s$ is the initial spreader, 
 then the probability $P^{(t)}_{ \to i}$ that node $i$ receives the infection, from an arbitrary initial spreader, at exactly stage $t$ of the dynamics is estimated as
\begin{equation}
P^{(t)}_{ \to i}= \sum_{s=1}^N \, z_{s} \,  P_{s\to i}^{(t)} \; .
\label{eq:P_multi}
\end{equation}
Similarly, the probability $Q^{(t)}_{\to i}$ that node $i$ receives the infection by an arbitrary single source at time $t$ or earlier is given by
\begin{equation}
    Q^{(t)}_{\to i}= \sum_{s=1}^N  z_s \, Q_{s \to i}^{(t)} \; .
    \label{eq:Q_multi}
\end{equation}

\subsubsection*{\label{sec:uncond}Individual-node mean-field approximation}

The individual-based mean-field approximation (IBMFA) consists in neglecting dynamical correlations among variables so that 
every node $i$ feels the average, over an infinite number of independent realizations of the spreading process, behavior of its neighbors~\cite{wang2003proceedings,Chakrabarti_2008, pastor2015epidemic}. Under IBMFA, we can write
\begin{equation}
    P_{s \to i}^{(t)} = \left[1 - Q_{s \to i}^{(t-1)}\right] \, \left[1 - \prod_{j=1}^N \left(1 -  A_{ji} \, \beta \, Q_{s \to j}^{(t-1)}\right)\right] \; .
    \label{eq:IBMFA_P}
\end{equation}
Eq.~(\ref{eq:IBMFA_P}) is derived as follows. The probability for node $i$ to be infected at exactly time $t$ is given by the product that the node has been not infected at any time earlier than $t$, i.e.,  $1 - Q_{s \to i}^{(t-1)}$, and receives the infection by at least one of its infected neighbors, i.e., $1 - \prod_{j=1}^N \left(1 -  A_{ji} \, \beta \, Q_{s \to j}^{(t-1)}\right)$. The latter term is computed as the product of individual contributions of the node's neighbors, thus assuming complete independence of their states. 

Eq.~(\ref{eq:IBMFA_P}), together with Eq.~(\ref{eq:Q_sing}), defines a system of $N$ equations, one  for every node $i$. Solutions  
are obtained by iteration, starting from the imposed initial conditions. 

Limitations of IBMFA are apparent. 
Neglecting dynamical correlations leads to the possibility for the infection to spread in opposite directions along the same edge, a situation that is indeed impossible in SI dynamics. 
Because of this fact, the cumulative probability $Q_{s \to i}^{(t)}$ always overestimates the true probability value, thus providing a consistent upper bound for the ground truth. Approximations more precise than IBMFA can be obtained by accounting for dynamical correlations tracing back the evolution of the system over a given number of time steps, considering additional variables representing the states of pairs, triplets, etc. of nodes~\cite{eames2002modeling,gleeson2011high}. Better approximations require fully accounting for the unidirectional motion of the infection along network edges.
For example, 
Lokhov and collaborators~\cite{lokhov2015dynamic} rely on conditional independence among variables, and write equations of messages spreading along individual edges, in the same spirit as done in approximations used in the study of percolation models on networks \cite{hamilton2014tight,karrer2014percolation,radicchi2015percolation,radicchi2015breaking,radicchi2018uncertainty}. 
Their dynamic message-passing approximation (DMPA) is exact on trees. In networks with loops, DMPA still leads to an overestimation of the true $Q_{s \to i}^{(t)}$, as the infection is allowed to travel in opposite directions on the same edge, although not immediately.

\subsubsection*{\label{sec:level1}Shortest-path combinatorial approximation}

The exact computation of $P_{s \to i}^{(t)}$ requires the enumeration of all possible ways in which the infection starting from node $s$ reaches node $i$ in exactly $t$ time steps. Such an enumeration includes all possible paths among the two nodes, and all possible combinations for the propagation of the infection along these paths. For an arbitrary network, the number of possibilities grows exponentially with the system size, thus making the exact computation of $P_{s \to i}^{(t)}$ infeasible. Here, we propose a way to approximate from below $P_{s \to i}^{(t)}$ by simply assuming that the spread of the infection may happen only along the shortest path connecting the nodes $s$ and $i$, and then enumerating all possible ways in which the infection can propagate along such a path. We name the approximation as the shortest-path combinatorial approximation (SPCA). We stress that SPCA is exact on trees, where each pair of nodes is connected by a unique path. We expect SPCA to provide a tight lower bound for the ground-truth value of $P_{s \to i}^{(t)}$ in sparse loopy networks. Under SPCA, $P_{s\to i}^{(t)}$ is uniquely determined by the distance $\ell_{si}>0$ between nodes $s$ and $i$. We can write
\begin{equation}
P_{s\to i}^{(t)}=\binom{t-1}{\ell_{si}-1} \, \beta^{\ell_{si}} \, (1-\beta)^{t-\ell_{si}} \;. 
\label{eq:P_sing}
\end{equation}
\begin{figure}[!hbt]
    \begin{center}
    \includegraphics[width=.45\textwidth]{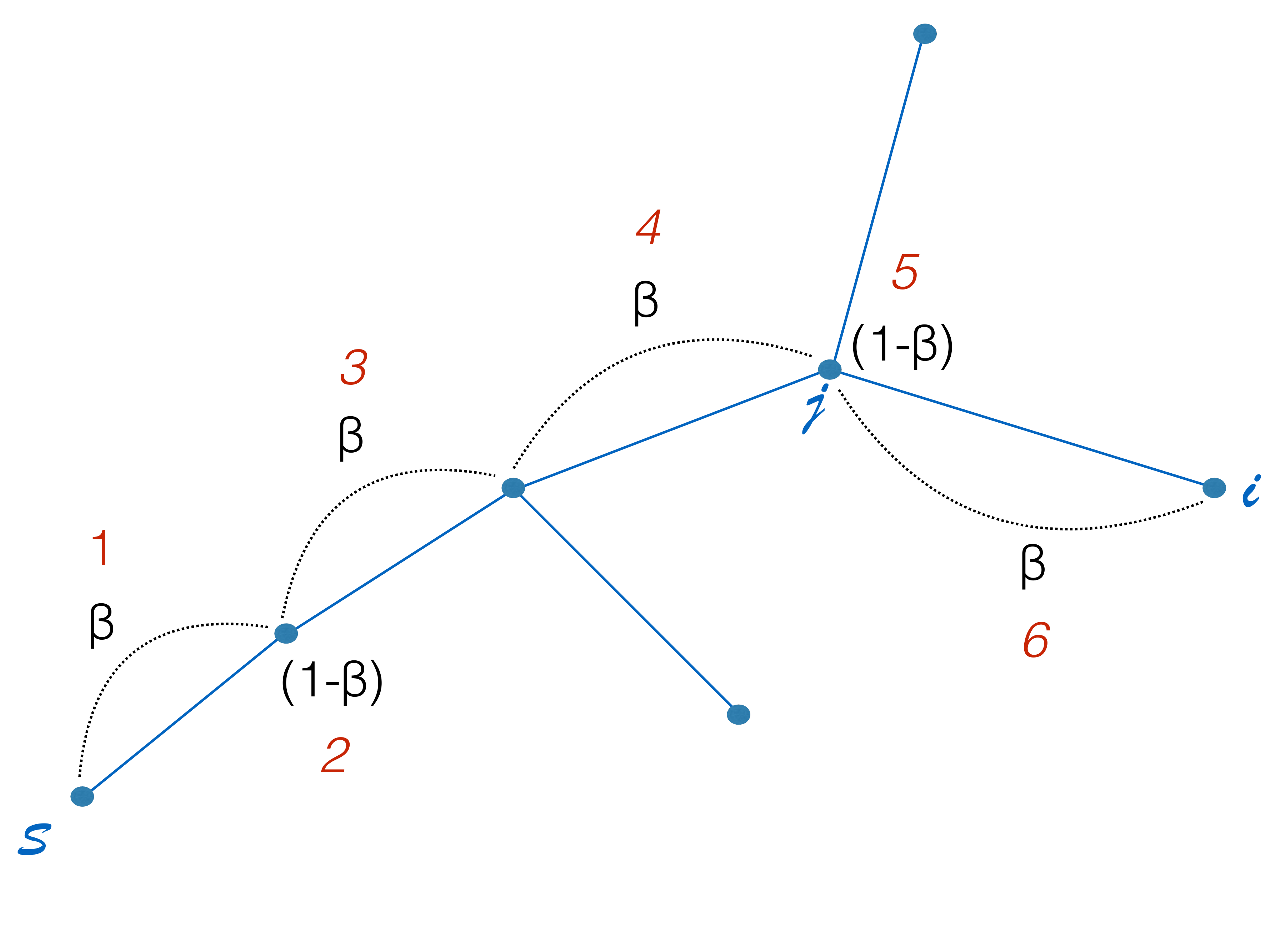}
    \end{center}
    \caption{The shortest-path combinatorial approximation. The figure serves to illustrate the rationale behind Eq.~(\ref{eq:P_sing}). Here, we represent a specific sequence of spreading attempts that allow the infection to spread from the source node $s$ to node $i$ at distance $\ell_{si}=4$ in exactly $t=6$ time dynamical steps. The sequence consists of four successful spreading events, and two unsuccessful spreading attempts, thus the probability of the sequence is $\beta^4 (1- \beta)^2$.}
    \label{fig:Path_prob}
\end{figure}

Eq.~(\ref{eq:P_sing}) is derived as follows. 
We can think \change{of} the path $s \to i$ as composed of two pieces, $s \to j$ and $j \to i$, where $j$ is the nearest neighbor of node $i$ along the path $s \to i$, so that the distance between $s$ and $j$ is $\ell_{sj} = \ell_{si} - 1$, as in Figure~\ref{fig:Path_prob}.
At stage $t$ of the dynamics, the final spreading attempt $j \to i$ must be, by definition of $P_{s \to i}^{(t)}$, successful. This elementary event happens with probability $\beta$. However before the final step, the infection must have reached node $j$ and not moved further than node $j$ in the preceding $t-1$  steps of the dynamics. This fact happens with  probability equal to $\binom{t-1}{\ell_{si}-1} \, \beta^{\ell_{si}-1} \, (1-\beta)^{t-\ell_{si}}$, corresponding to the binomial probability of observing exactly $\ell_{si}-1$ successful spreading events in $t-1$ total spreading attempts. Eq.~(\ref{eq:P_sing}) is finally given by the product of above-defined probabilities for the two pieces $s \to j$ and $j \to i$ of the path $s \to i$.

$Q_{s \to i}^{(t)}$ is obtained relying on Eq.~(\ref{eq:Q_sing}). It is easy to check that $\lim_{t\to\infty} Q_{s\to i}^{(t)}=1$, 
for all nodes $i$ and for any source node $s$, as expected for the SI model. Regardless of its distance from the source, every node will be eventually infected.

\begin{figure}[!hbt]

\includegraphics[width=0.45\textwidth]{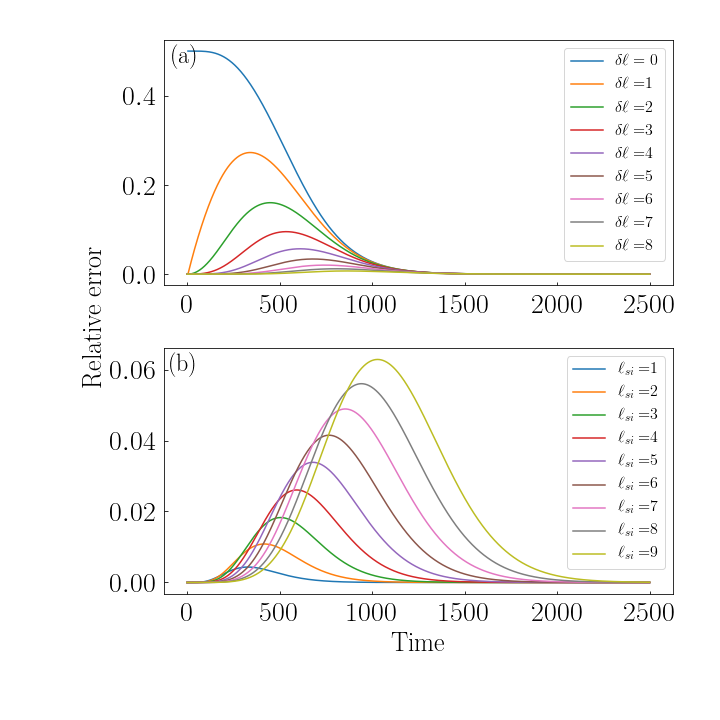}
\caption{Error committed by the shortest-path combinatorial approximation.
Relative error of the SPCA compared to the ground-truth value for the probability $Q_{s \to i}^{(t)}$ that node $i$ is infected at time $t$ or earlier by an infection starting from node $s$. Spreading obeys SI dynamics with spreading probability  $\beta=0.01$. Ground-truth values are estimated under the hypothesis that nodes $s$ and $i$ are connected by two independent paths of length $\ell_{si}$ and $\ell_{si} + d \ell$, respectively. SPCA approximates truth probabilities relying on the shortest path only. (a) We set $\ell_{si} = 5$ and consider different values of $d \ell$. Relative error is plotted as a function of time. (b) We set $d \ell = 5$, and consider different $\ell_{si}$ values.
}
\label{fig:eps_si}
\end{figure}

To get a sense of the magnitude of the approximation error introduced by SPCA, we consider the case where nodes $s$ and $i$ are connected by two independent paths of length $\ell_{si}$ and $\ell_{si} + d \ell$, respectively. We focus our attention on the probability $Q^{(t)}_{s \to i}$ that the infection reaches node $i$ at time $t$ or earlier. Such a probability is given by the likelihood that infection propagates along at least one of the two independent paths, and its ground-truth value can be calculated exactly relying on a proper combination of Eqs.~(\ref{eq:Q_sing}) and~(\ref{eq:P_sing}), see appendix for details. We compare the ground-truth value with the one we obtain using SPCA, and quantify the relative error of the approximation with respect to the truth value. Results for some combinations of the parameter values $\ell_{si}$ and $d \ell$ are plotted in Figure~\ref{fig:eps_si}. Relative error is a decreasing function of $d \ell$. It is worth noting that the relative error behaves non-monotonically as a function of time $t$, reaching a maximum at intermediate $t$ values.

\subsubsection*{Predictions in real-world networks}

\begin{figure}[!hbt]

\includegraphics[width=0.45\textwidth]{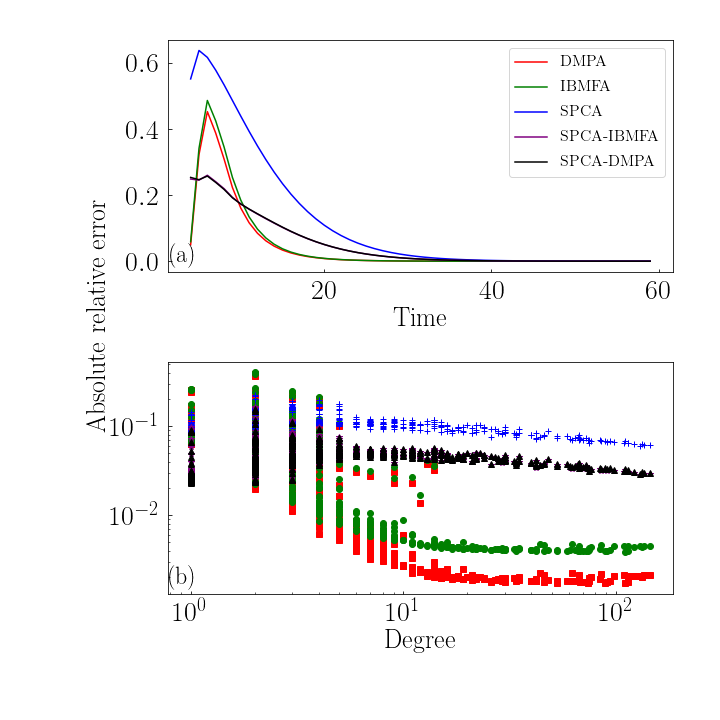}
\caption{Accuracy of the approximations in predicting ground-truth infection probabilities in real-world networks. We considered the SI model on the air transportation network of Ref.~\cite{colizza2007reaction}. The spreading probability is set  $\beta = 0.25$. We run $4,000$ numerical simulations of the process where a given node is the source of the spreading. 
Ground-truth values are compared with predictions from the IBMFA (green), SPCA (blue), DMPA (red), the average SPCA-IBMFA (purple) and SPCA-DMPA (black). a) Mean absolute error over all possible sources as a function of time. b) We display the absolute error averaged over time as a function of the degree of the source node.
}
\label{fig:SI_real_net}
\end{figure}

A nice property of SPCA is to provide a consistent lower-bound for ground-truth probabilities. SPCA neglects that spreading may occur along longer paths. The simultaneous use of IBMFA 
(or similar approaches that provide upper bounds for true probabilities, e.g., DMPA) 
and SPCA is very useful, as it allows us to delineate the region of possible outcomes for the SI model. In Figure~\ref{fig:SI_real_net} for example, we compare predictions of IBMFA, DMPA and SPCA with estimates of the ground-truth probabilities obtained via numerical simulations of the SI process on a real network. \change{We use the US air transportation network of Ref.~\cite{colizza2007reaction}. The network has $N=500$ nodes, density of connections $\rho \simeq 0.024$,  and diameter $d=7$.} 
Results are averages from random initial placements of the source of the infection, thus, in the various theoretical approximations, the prior of Eq.~(\ref{eq:prior}) is taken equal to the uniform distribution, i.e.,  $z_s = 1/N$.   

While taking into account only the shortest path is an oversimplification of the problem, combining SPCA with IBMFA (or DMPA), as for example by taking the arithmetic average of the two approximations, can increase the accuracy of the individual methods. This is especially true for sources of the process with low degree.

\subsection{\label{sec:sir}The susceptible-infected-recovered model}

\subsubsection*{Problem setting}

For SIR dynamics, we consider the same initial configuration as in the SI model, where spreading is initiated by a single infected node $s$, i.e., $\sigma_{s}^{(0)} = I$, while all other nodes $j$ are initially in the susceptible state, i.e., $\sigma_{\forall j \neq s}^{(0)} = S$. The probability that node $i$ receives the infection at exactly time $t$ is defined in the same identical way as for the SI model, see Eq.~(\ref{eq:P}). We can further apply the same definition as in Eq.~(\ref{eq:Q_sing}) to quantify the probability $Q_{s \to i}^{(t)}$ that the infection reaches node $i$ at time $t$ or earlier. The additional recovered state that is allowed in the SIR model requires the definition of other probabilities not defined for the SI model. For instance, the probability that node $i$ recovers exactly at time $t$, given that the infection started from node $s$, is denoted by $R_{s \to i}^{(t)}$.
 The probability that node $i$ recovers at time $t$ or earlier is given by 
\begin{equation}
T_{s \to i}^{(t)}= \sum_{r=0}^{t}  R_{s\to i}^{(r)} \,  \; .
\label{eq:T}
\end{equation}
All other probabilities of interest can be immediately derived. For example, we have that the probability that node $i$ is still in the susceptible state at time $t$ is 
given by $1- Q_{s\to i}^{(t)}$. Also, the probability that node $i$ is found infected at time $t$ is given by $Q_{s\to i}^{(t)} - T_{s \to i}^{(t)}$. 

\subsubsection*{\label{sec:IBMFA_sir}Individual-based mean-field approximation}

Under IBMFA~\cite{wang2003proceedings,Chakrabarti_2008, pastor2015epidemic}, we can write
\begin{equation}
P_{s \to i}^{(t)} =  \left[1 - Q_{s \to i}^{(t-1)}\right] \, 
 \left\{1 - \prod_{j=1}^N \left[ 1 -  A_{ji} \, \beta \, 
 \left(Q_{s\to j}^{(t-1)} - T_{s \to j}^{(t-1)}
 \right) 
 \right]\right\} 
\; .
    \label{eq:SIR_IBMFA}
\end{equation}
Eq.~(\ref{eq:SIR_IBMFA}) is a direct generalization of Eq.~(\ref{eq:IBMFA_P}). The meaning of the various terms is exactly the same as in Eq.~(\ref{eq:IBMFA_P}), with the only difference that here we need to account for the possibility of infected nodes to recover. To become infected, we require that node $i$ is still in the susceptible state at time $t$, i.e., $1 - Q_{s \to i}^{(t-1)}$, and that infection arrives from at least one of its neighbors that is still in the infected state, i.e., 
$1 - \prod_{j=1}^N \left[ 1 -  A_{ji} \, \beta \, \left(Q_{s\to j}^{(t-1)} - T_{s \to j}^{(t-1)}\right) \right]$.

Properties and limitations of IBMFA for the SIR model are very similar to those already illustrated for the SI model. Limitations of Eq.~(\ref{eq:SIR_IBMFA}) in capturing the true probability are due to the assumption of dynamical independence among variables that is in contrast with the unidirectional nature of spreading. Still, IBMFA can be improved by imposing only conditional independence instead of full independence among variables as done in DMPA~\cite{lokhov2015dynamic}. Also, IBMFA and DMPA continue to provide effective methods to bound from the above marginal probabilities of infection of the true SIR model.

\subsubsection*{\label{sec:tbca_sir}Shortest-path combinatorial approximation}

In this section, 
we extend SPCA to the SIR model.  
The probability that node $i$ is infected 
exactly at time $t$, given that the infection started from the source node $s$, is
\begin{equation}
P_{s\to i}^{(t)}= (1-\gamma)^{t-\ell_{si}} \, \binom{t-1}{\ell_{si}-1} \, \beta^{\ell_{si}} \, (1-\beta)^{t-\ell_{si}} \; . 
\label{eq:sir_P_sing}
\end{equation}
Eq.~(\ref{eq:sir_P_sing}) simply generalizes Eq.~(\ref{eq:P_sing}) with the inclusion of the multiplicative factor $(1-\gamma)^{t-\ell_{si}}$. This factor accounts for eventual recovery events that may prevent the infection to reach node $i$. As the infection can proceed its trajectory as long as the latest infected node does not recover before passing the infection, then the condition that allows spreading to occur is that recovery should not happen in $t-\ell_{si}$ independent attempts, leading to the factor $(1-\gamma)^{t-\ell_{si}}$.

We can immediately derive that
\begin{equation}
\lim_{t \to \infty} Q_{s\to i}^{(t)}= \left( 1-\gamma+\frac{\gamma}{\beta} \right)^{-\ell_{si}} \; ,
\label{eq:lim_inf}
\end{equation}
thus, in the long-term limit, the probability of infection exponentially decreases towards zero 
as the distance from the source increases. 
 
\begin{figure}[!hbt]
\centering

\includegraphics[width=0.45\textwidth]{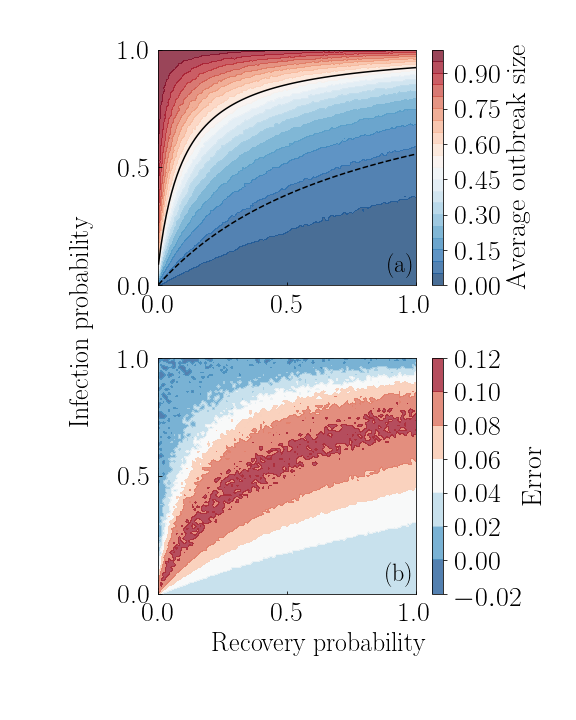}

\caption{Prediction of the phase diagram under the shortest-path combinatorial approximation.
a) We consider a tree with $N=50$ nodes and uniformly distributed random Pr\"ufer sequence~\cite{prufer1918neuer, pemmaraju2003computational}. \change{We display the ground-truth value of the outbreak size, estimated from numerical simulations, as a function of the spreading and recovery probabilities. The two black curves correspond to the solutions of Eq.~(\ref{eq:avg_lim}) obtained by setting respectively $O = 0.5$ (full curve) and $O = 0.01$ (dashed). 
b) Difference between ground-truth values of the outbreak size and predictions from Eq.~(\ref{eq:avg_lim}) as a function of the SIR model parameters.}
}
\label{fig:outbreak}
\end{figure}

In spite of the fact that Eq.~(\ref{eq:lim_inf}) is valid for individual nodes only, 
we find that the equation is useful for the prediction of the outbreak of the entire system. 
In Figure~\ref{fig:outbreak}, we compare the average value of the outbreak size estimated from numerical simulations
with predictions quantified as
\begin{equation}
O = \left( 1-\gamma+\frac{\gamma}{\beta}\right)^{- \langle \ell \rangle} \; ,
\label{eq:avg_lim}
\end{equation}
where $\langle \ell \rangle = \frac{2}{N(N-1)} \, \sum_{i>j} \ell_{i,j}$ is the average value of the distance among nodes in the network. While predictions do not match truth values, their functional similarity is apparent as shown  by the magnitude of the mismatch between ground-truth values and predictions.

\begin{figure}[!hbt]
\centering
\includegraphics[width=0.5\textwidth]{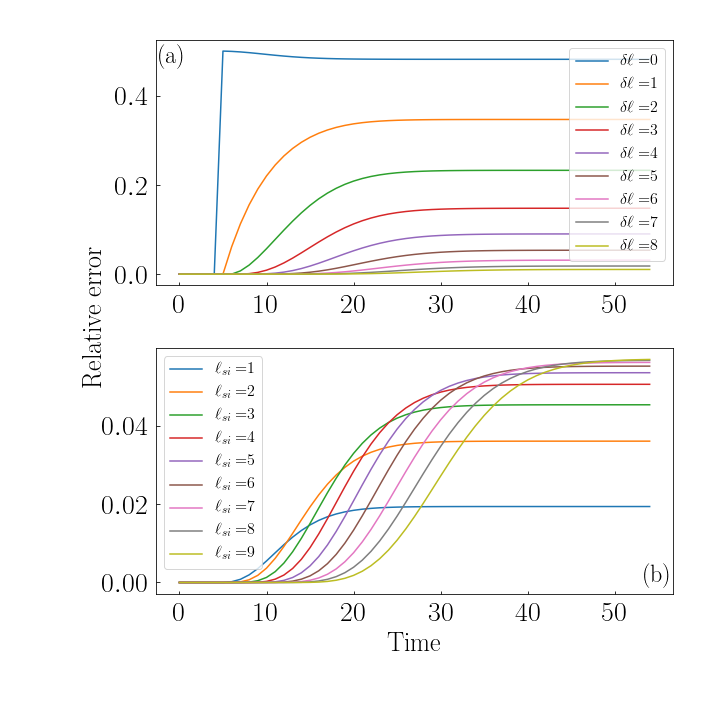}

\caption{Error committed by the shortest-path combinatorial approximation.
Same as in Figure~\ref{fig:eps_si} but for the SIR model. 
Spreading probability is $\beta=0.5$, while recovery probability is $\gamma=0.5$.}
\label{fig:eps_sir}
\end{figure}

The probability to recover at  time $t$ is given by
\begin{equation}
R_{s \to i}^{(t)}=\gamma \, (1-\gamma)^{t-1} \, \sum_{r=0}^{t-1}  P_{s\to i}^{(r)} \, (1-\gamma)^{-r} \; .
\label{eq:R}
\end{equation}
Eq.~(\ref{eq:R}) is easily obtained considering that node $i$ can recover only if previously infected, say at time $r$. Then, the probability that recovery happens after a certain number of additional time steps is given by the probability that recovery happened at time $t$ but didn't happen in any of the previous stages, i.e., $\gamma \, (1-\gamma)^{t-r-1}$. Summing up over all possible time steps $r$ when node $i$ could have been infected, one obtains Eq.~(\ref{eq:R}).

In Figure~\ref{fig:eps_sir}, we repeat the same exercise as in Figure~\ref{fig:eps_si} by estimating the relative error  committed by SPCA when the ground-truth topology is such that nodes $s$ and $i$ are connected by two independent paths of length $\ell_{si}$ and $\ell_{si} + d \ell$, respectively. We note that the behaviour in the early stages of the relative error is quite similar to the SI case. Results for the SIR model differ from those of the SI model in the late stages of the dynamics, when the finite limit in Eq.~(\ref{eq:lim_inf}) gives a non-vanishing asymptotic value. For the SIR model, the eventual presence of multiple paths connecting two nodes plays a much more important role than in the SI model.

\subsubsection*{Predictions in real-world networks}

\begin{figure*}[!hbt]

\includegraphics[width=0.95\textwidth]{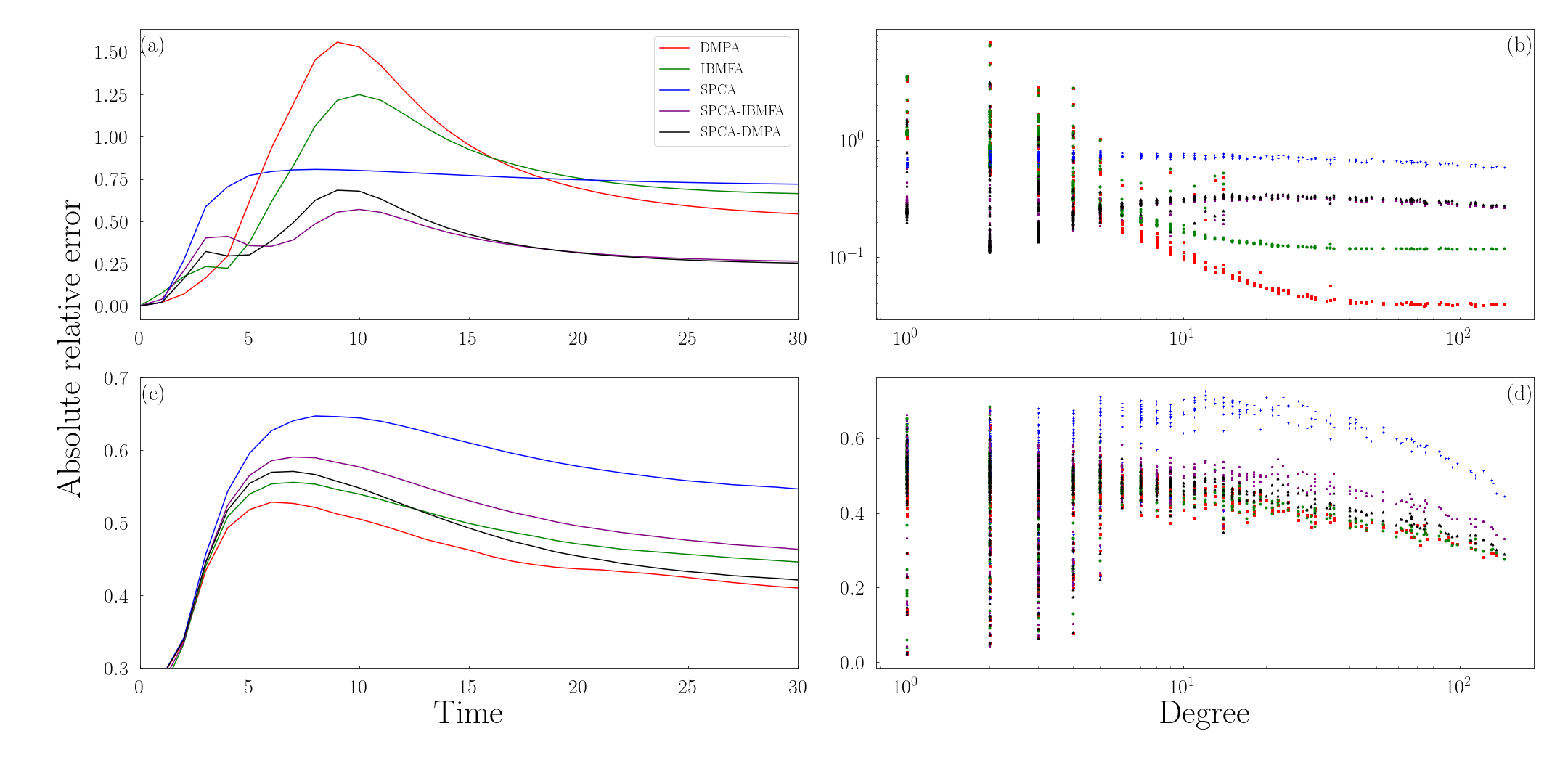}
\caption{Accuracy of the approximations in predicting ground-truth infection probabilities in real-world networks. a) We consider the SIR model on the air transportation network of Ref.~\cite{colizza2007reaction}. We set $\beta = 0.1$ and $\gamma = 0.1$. The combination of the two parameter values correspond to the supercritical regime of the dynamics. We run $8,000$ numerical simulations of the SIR process for each node being the source of the infection to obtain a single estimate of  the ground-truth values of $Q_{\to i}^{(t)}$ for all $i$. Ground-truth values are compared with predictions from the IBMFA (green), SPCA (blue), DMPA (red), the average SPCA-IBMFA (purple) and the average SPCA-DMPA (black). The figure displays the relative error, averaged over all nodes, committed by the various approximations as a function of time. b) Absolute error, averaged over time,  of the various approximations as a function of the degree of the source node. Data are the same as in panel a. c) Same as in panel a, but for $\beta = 0.01$, corresponding to the subcritical  regime of spreading. d) Same as in panel b, but obtained for the same parameter setting as in panel c.
}
\label{fig:SIR_real_net}
\end{figure*}

The considerations made for the SI model are still valid for the SIR model. In networks with loops, SPCA underestimates the ground-truth probability $Q_{s \to i}^{(t)}$, while DMPA overestimates it. The combination of SPCA and DMPA defines the region where the true values are located. 
Also, it is possible to improve the accuracy of the individual approximations by simply taking their arithmetic average, see Figure~\ref{fig:SIR_real_net}. Improvements are especially apparent in the supercritical regime of the dynamics, where the SIR model behaves most similarly to the SI model.

\section{\label{sec:applications}Applications}

We now turn our attention to specific applications of the theoretical framework in 
inference problems. We remark that we are not leveraging the framework to actually perform inference. Rather, we are using it to provide insights on the properties of the inference problems, as for example how the ability of an observer to perform inference is affected by the time of the observations and the position of the observer in the system. 

\subsection{\label{sec:si_patient}Identification of the source of spreading}
As a first application, we consider the classical inference problem aiming at the identification of the initial spreader, i.e., the so-called patient-zero problem \cite{shah2010detecting,shah2011rumors,luo2013identifying}. We note that the problem has been already studied with a combinatorial approach similar to ours in Ref.~\cite{zhu2014information}. The patient-zero problem is typically framed under the assumption of limited information, where only partial knowledge of the microscopic properties of the system is available to the observer. In this paper, we  focus on two different settings often considered in the literature on this subject. In both cases, we assume that the observer has full and exact knowledge of the network topology. In addition, the observer is fully aware of the stage of the dynamics as well as of the exact values of the  spreading and recovery probabilities.

\subsubsection*{Susceptible-infected model}
First, we consider the case where the
observer is allowed to constantly monitor the state of 
node $i$. Suppose that, at time $t$, node $i$ gets infected, and the observer wants to infer the identity of the patient zero. The probability $V_{s \to i}^{(t)}$ that node $s$ is the initial spreader is given by the Bayes' theorem and can be written as
\begin{equation}
V_{s \to i}^{(t)}= \frac{z_s \, P_{s \to i}^{(t)}}{P_{ \to i}^{(t)}} \; ,
\label{eq:P_zero}
\end{equation}
where $P_{s \to i}^{(t)}$ and $P_{ \to i}^{(t)}$ are the same probabilities as defined in Eqs.~(\ref{eq:P_multi}) and~(\ref{eq:P_sing}), respectively. $z_s$, defined in Eq.~(\ref{eq:prior}), is the probability that node $s$ is the source of the infection prior any type of measurement made on the system. 

Second, we consider the  case where the observer performs a single measurement of node $i$ at time $t$ and finds it infected. Infection may have occurred at time $t$ or earlier. The probability $W_{s \to i}^{(t)}$ that node $s$ is the source of the infection under this condition is given by
\begin{equation}
W_{s \to i}^{(t)}=\frac{z_s \, Q^{(t)}_{s \to i}}{Q^{(t)}_{ \to i}}  \; .
\label{eq:Q_zero}
\end{equation}
Here, $z_s$ is still our prior on the node $s$ acting as the source of the infection; $Q_{s \to i}^{(t)}$ and $Q_{ \to i}^{(t)}$ are the same probabilities as defined in Eqs.~(\ref{eq:Q_sing}) and~(\ref{eq:Q_multi}), respectively.

\begin{figure}[!hbt]
\centering

\includegraphics[width=8cm]{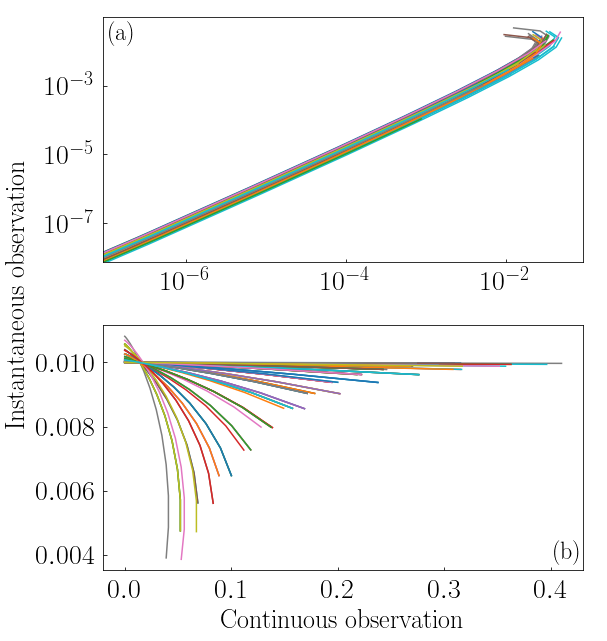}
\caption{Comparison of observation strategies in the patient-zero identification problem. We display 
the inferred probabilities on the location of \change{the source $s$} obtained under the hypothesis of continuous observation of the state of nodes, i.e., $V_{s \to i}^{(t)}$ as defined in Eq.~(\ref{eq:P_zero}), 
and under the hypothesis of instantaneous observation of their state, i.e., $W_{s \to i}^{(t)}$ as defined in Eq.~(\ref{eq:Q_zero}). \change{Each curve corresponds to a single measured node $i$ at time $t$; the curve is obtained by connecting pairs of contiguous points ($V_{s \to i}^{(t)},W_{s \to i}^{(t)}$) for all $s \neq i$}.  The two panels show results valid for \change{two different values of the time of measurement}, namely $t=12$ in panel a and $t=90$ in panel b. Results are obtained on a tree  with $N=100$ nodes and uniformly distributed random Pr\"ufer sequence~\cite{prufer1918neuer, pemmaraju2003computational}.
Spreading is happening according to the SI model with spreading probability $\beta =0.3$.}
\label{fig:pst}
\end{figure}

A comparison between  $V_{s \to i}^{(t)}$ and $W_{s \to i}^{(t)}$ is illustrated in Figure~\ref{fig:pst}. The
two strategies of performing local observations of the network generally lead to different predictions about the location of the patient zero, and the difference among the two strategies strongly depends on the time when the measurements are performed.

In the early stages of the spreading process, the values of $V_{s \to i}^{(t)}$ or $W_{s \to i}^{(t)}$ are highly heterogeneous, and the two strategies of observation lead to almost identical inferred probabilities regardless of the point of observation $i$. This fact is easily explained as the most likely source of infection should be located in the vicinity of the node where the system is observed from. At later stages of the dynamics, the probability values $V_{s \to i}^{(t)}$ or $W_{s \to i}^{(t)}$ are less heterogeneous, the two inferred probabilities are negatively correlated, and their discrepancy strongly depends on the node $i$ where the system is observed from. The negative correlation in the final stages of the dynamics seems surprising but can be intuitively explained. If a node gets infected after a very long time, it is very unlikely that the infection started from one of its neighbors. However, if one finds the node infected but does not know when the infection happened, still the nearest nodes are the most likely sources of infection.

\begin{figure}[!hbt]
    \centering
    \includegraphics[width=.45\textwidth]{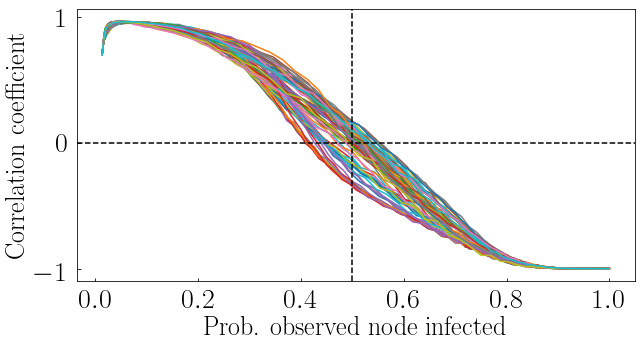}
    \caption{Comparison of observation strategies in the patient-zero identification problem.
    Spearman correlation coefficient obtained by ranking nodes according to the inferred probabilities of being the source of spreading according to instantaneous and continuous observation of the state of a node. The correlation coefficient is plotted against the probability  to find the observed node $i$ infected. Different curves correspond to different nodes observed, and different values of the probability to find the observed node infected map to different stages of the dynamics. Results are obtained in the same experimental setting as of Figure~\ref{fig:pst}.}
    \label{fig:RhoS}
\end{figure}

The quantity that best characterizes the correlation between the inferred probabilities  $V_{s \to i}^{(t)}$ and $W_{s \to i}^{(t)}$ is $Q^{(t)}_{\to i}$, i.e., the probability to find node $i$ infected at time $t$ or earlier. In Figure~\ref{fig:RhoS}, we display the Spearman correlation coefficient between $V_{s \to i}^{(t)}$ and $W_{s \to i}^{(t)}$ as a function of $Q^{(t)}_{\to i}$. Correlation is only mildly dependent on the node where the system is observed from. The non-perfect correlation found at very low values of $Q^{(t)}_{\to i}$ is still surprising, but can be easily understood. If one finds the observed node infected at the very first stages of the dynamics without knowing exactly the time when the node was  infected, then it is very likely that the node itself is inferred to be the patient zero. However, when one knows that the exact time of infection, one can properly distinguish cases when the observed node was or was not the actual source of spreading. There are two remarkable aspects emerging from Figure~\ref{fig:RhoS}. First, the change in sign of the correlation coefficient is happening at $Q^{(t)}_{\to i} \simeq 0.5$. Second, the correlation coefficient becomes maximally negative well before $Q^{(t)}_{\to i} \simeq 1$. This is due the fact that, at very late stages of the spreading when all nodes are likely to be found infected, having knowledge that the infection happened at a very late stage makes likely that the source of infection was far apart from the observation point; not having knowledge of the exact time of infection leads instead to an almost flat probability distribution for the patient zero location but still with a very weak preference for nodes close to the observation point, including the observed node itself. 

\subsubsection*{Susceptible-infected-recovered model}

In the SIR model, three  outcomes are possible when the state of a node is measured. As a consequence,  four different conditional probabilities are potentially relevant for the patient-zero identification problem. However,  finding the observed node in the infected vs. recovered state does not generate a significant difference in our ability to predict the identify of the patient zero.  

In Figure~\ref{fig:RhoS}, we measure the correlation between the inferred probabilities $V_{s \to i}^{(t)}$ and $W_{s \to i}^{(t)}$ on the location of the patient zero. We notice that the two strategies of observation may lead to very different outcomes depending on either the time when the observation is made and the values of the parameters of the spreading model. Clearly, for $\gamma=0$ we recover the same results as of the SI model, with correlation never increasing as a function of time. For values of the recovery probability $\gamma > 0$, correlation is instead a non-monotonic function of time. At early stages of the dynamics, correlation decreases for the same reasons as in the SI model. At late stages, however, it increases. The reason is quite intuitive. Finding a node infected but not yet recovered at a late stage of the dynamics means that it is unlikely that the node got infected at the very beginning process, otherwise it would have had plenty of time to recover. Thus, knowing or not knowing the exact time of infection is irrelevant for the patient-zero inference problem, as one can exclude that the source of infection is very close to the observed node in either cases.

\begin{figure}[!hbt]
    \centering
    \includegraphics[width=.45\textwidth]{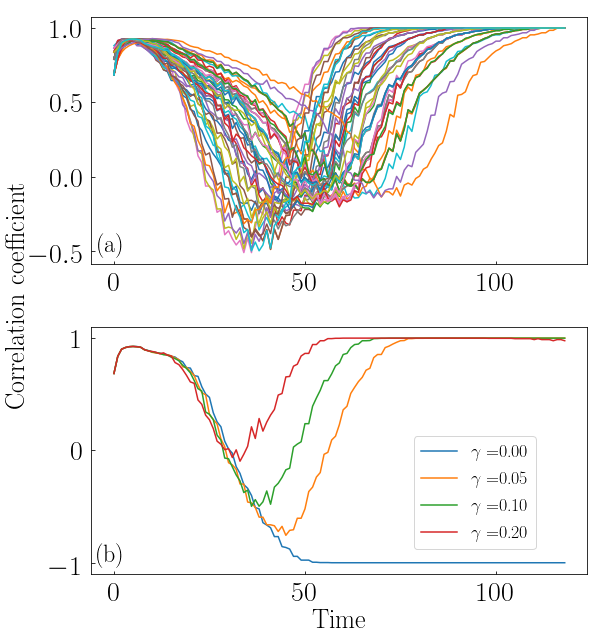}
    \caption{Comparison of observation strategies in the patient-zero identification problem in the SIR model. a) Spearman correlation coefficient obtained by ranking nodes according to the inferred probabilities of being the source of spreading according to instantaneous and continuous observation of the state of a node. The correlation coefficient is plotted against time. Different curves correspond to different nodes observed. Results are obtained in the same configuration as of Figure~\ref{fig:pst}. Spreading probability is set $\beta=0.2$, while recovery probability is set $\gamma=0.1$. b) Same as in panel a, but for different $\gamma$ values. Irrespective of the specific value of the recovery probability, the system is always observed from the same node. 
    }
    \label{fig:RhoSIR}
\end{figure}

\subsection{\label{sec:gain}Gain of information from local measurements}

What is the amount of information, about the microscopic configuration of the system, that we can gain by observing the state of a specific node? Clearly, as states of different nodes in the network are correlated, the measurement of the state of one node provides us with some knowledge about the state of the other nodes. However, the gain of information will be not the same for all choices of the observed node; further, the gain of information may dramatically vary, even if we decide to observe the same node, depending on the stage of the dynamics when the measurement is performed. 

As a second application of our framework, we study spreading processes from a information-theoretical 
perspective providing indications about the content of information that each node carries about the whole network.

To properly quantify the gain of information we should calculate the mutual information between network configurations and the state of the observed node. This calculation would require to estimate the probability of every network configuration conditioned by the state of the observed node, and then a sum over all the possible conditional probabilities. Due to the huge number of possible configurations however, the exact computation of the information gain is infeasible. Here, we approximate it as the sum of the pairwise mutual information of all pairs of nodes. We compute the mutual information among pairs of nodes $i$ and $j$ using their joint probability of getting infected at time $t$ or earlier, namely  $Q^{(t)}_{s\to i,j}$. The geometric framework of SPCA easily adapts to such a computation. Specifically,  the computation of the joint probability still relies on the definition of marginal probabilities, but properly accounts for the possible paths between the source $s$ and the two target nodes $i$ and $j$ we are interested in (see Appendix for details). The expected information gained by observing node $i$ is then quantified as the sum of the pairwise mutual information of the node with respect to all other nodes in the network.

\subsubsection*{Susceptible-infected model}
\begin{figure}[!hbt]
\centering
\includegraphics[width=0.45\textwidth]{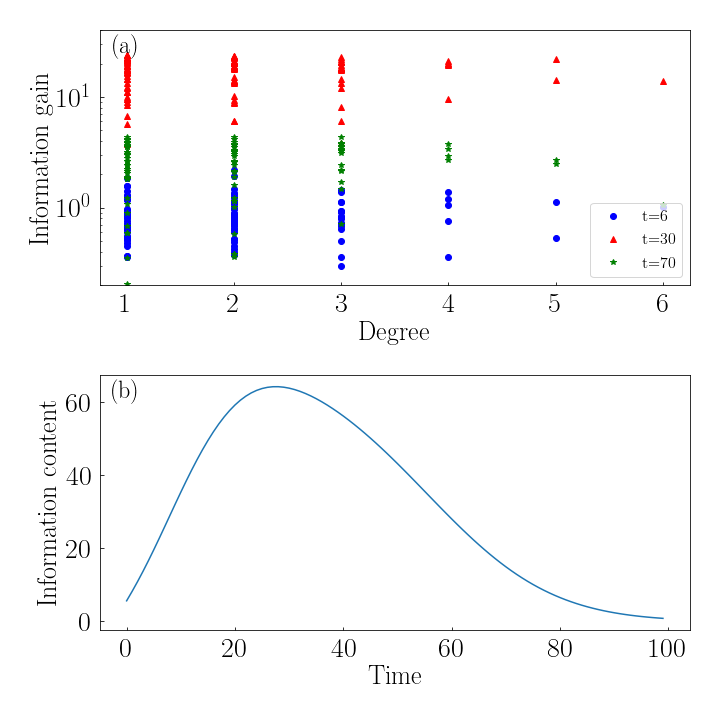}
\caption{Gain of information from local network measurements in the SI model. a) Gain of information obtained from the observation of a single node. Information gain is plotted against the degree of the observed node. Each point in the plot corresponds to a different node used to observe the system. Different colors and symbols stand for different stages of the dynamical process when the observation is performed. The experimental setting is the same as in Figure~\ref{fig:pst}. b) Total information content of the network as a function of time. Information content is measured as the sum of the individual-node entropies, i.e., $I = \sum_i [ Q_{\to i}^{(t)} \log Q_{\to i}^{(t)}+(1-Q_{\to i}^{(t)})\log(1-Q_{\to i}^{(t)})]$.}
\label{fig:IG}
\end{figure}

 Naively, we should expect that nodes that occupy central positions in the network correspond to optimal points of observations. Such an intuition is generally correct, but with some caveat. In Figure~\ref{fig:IG}, we show the information gained by measuring a single node as a function of its degree, i.e., a simple metric of network centrality. \change{We considered metrics of network centrality more complicated than degree, but the results of the analysis are qualitatively similar to those reported here.} In the early stages of the dynamics, the information gain is correlated with node degree. At late stages, observing the network from nodes with large degrees becomes sub-optimal.
 
 The time of the measurement plays a fundamental role for the amount of information that can be actually gained. At the beginning of the dynamics, all nodes are in the susceptible state, thus we do not expect any measurement to be informative. Similar conclusions are valid for the late stages of the dynamics, when all nodes are likely to be in the infected state. We expect, however, measurements to be  informative when uncertainty about the system configuration is maximal. This fact is apparent from  Figure~\ref{fig:IG}. We see that there is an intermediate stage of the dynamics where the information content of the network reaches a peak value. At that point in time, the information gained by observing a node is not strongly dependent on the centrality of the node where the observation is performed. 
 
 \subsubsection*{Susceptible-infected-recovered model}
\begin{figure}[!hbt]
\centering
\includegraphics[width=0.45\textwidth]{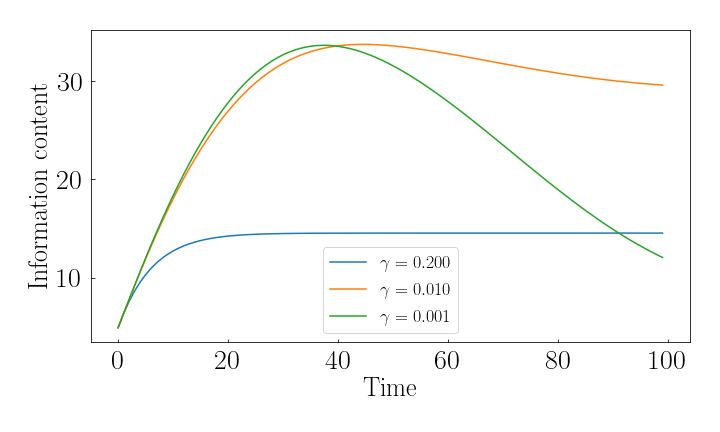}
\caption{Gain of information from local network measurements in the SIR model.  Total information content of the network as a function of time for three different values of $\gamma$. We set $\beta = 0.2$ and consider the  same network as in Figure~\ref{fig:RhoSIR}.}
\label{fig:IGSIR}
\end{figure}

 In Figure~\ref{fig:IGSIR}, we perform a similar analysis as of Figure~\ref{fig:IG}, but for the SIR model. The content of information as a function of time behaves in a different way depending on the choice of the parameters $\beta$ and $\gamma$. If the probability of recovery $\gamma$ is low, then the information content of the SIR model is almost identical to the one we just described for the SI model. If $\gamma$ values are large enough instead, information content does not longer decrease as time increases, reflecting the non-null uncertainty of the final configurations reached by SIR spreading.
 
\section{\label{sec:conclusions}Conclusions}

In this paper, we presented a  combinatorial approach to calculate the spreading probability, along the shortest path between pairs of nodes in a network, for the Susceptible-Infected (SI) and the Susceptible-Infected-Recovered (SIR) models. We named it as the shortest-path combinatorial approximation (SPCA). The approach is exact in absence of loops and gives a lower bound for the infection probability on arbitrary networks. The approximation can be in principle extended to include the effect of other paths (i.e., the second shortest path, the third shortest path, etc.) on the computation of the spreading probability. However, adding more paths exponentially increases the complexity of the algorithm, since the procedure requires to disentangle independent vs. shared parts (i.e., nodes and edges) among the various paths. We showed that the arithmetic average between the novel approximation and other approximations existing on the market, e.g., individual-node mean-field and dynamic message-passing approximations, can be used to obtain predictions that are more accurate than those obtained by each approximation if used in isolation. The amount of improvement strongly depends on the degree of the source and, in the SIR model, on the regime of the process, hinting that the importance of the shortest path depends on network's connectivity as well as on the process's parameters. Potential follow-up studies could explore the predictive power of more sophisticated ways than the arithmetic average to combine SPCA with other approximations. On a tree network, we used SPCA to evaluate joint probabilities among pairs of node states, and applied it to study general properties of standard inference problems. Specifically, we characterized two different strategies of single-node observation in the identification problem of the patient zero. We showed that the inference problem is highly sensitive to \change{the} modality in which the observation is performed. Measuring a node at a given time or monitoring it throughout the process may lead to opposite conclusions on the identity of the patient zero. Also, we analyzed the entropy of the processes and quantified the information gained, when the state of a node is measured, about the rest of the network. The most informative node is not the same throughout the entire process and the knowledge of the dynamical stage is crucial to optimize the information gained by a measurement. These results can be extended by considering the measurements, 
contemporaneous or sequential, of two nodes. By calculating a three-node joint probability one could measure the most informative pair of nodes and study different strategies for nodes' control.
While we focused only on the SI and SIR models, other spreading processes in discrete time can be studied with a similar theoretical approach.

\acknowledgements
DM and FR acknowledge support from the US Army Research Office (W911NF-16-1- 0104). FR acknowledges support from the National Science Foundation (CMMI-1552487).



\appendix

\section{Magnitude of the error associated with the shortest-path combinatorial approximation}

In Figures~\ref{fig:eps_si} and \ref{fig:eps_sir}, we
considered an hypothetical setting where the generic node $i$ is connected to the source node $s$ by two independent paths of length $\ell_{si}$ and $\ell_{si} + d \ell$, with $d \ell \geq 0$. The paths are independent in the sense that they do not share any node except for $s$ and $i$. This fact allows us to easily compute the exact probabilities for the ground-truth scenario by simply combining the probabilities of the individual paths. The setting is useful to understand the magnitude of the error that we should expect to have when using SPCA in a non-tree network, where multiple paths among nodes may exist. 
For simplicity of notation, but without loss of generality, we will use $\ell = \ell_{si}$ in the following description.

\subsection*{Susceptible-infected model}
For the SI model, the probability that the infection reaches a certain node along a path of length $\ell$ in $t$ time steps or less is given by
\[
q_{1} (\ell, t) = \sum_{r=0}^{t} \, \binom{t-1}{\ell-1} \, \beta^{\ell} \, (1-\beta)^{t-\ell}  \; , 
\]
The previous expression is nothing more than a mere combination of Eqs.~(\ref{eq:Q_sing})
and~(\ref{eq:P_sing}) of the main text. We just avoided to write an explicit dependence on the source and target nodes to simplify the expression. 
In presence of two independent paths, the probability that the infection reaches the target node is given by
\[
  q_{2}(\ell, \ell + d \ell, t) = 1 - [1 - q_{1} (\ell, t)] [1 - q_{1} (\ell + d \ell, t)] \; ,
\]
thus equal to the probability that spreading occurs at least on 
one of the two independent paths. 
The relative error of Figure~\ref{fig:eps_si} is finally quantified as
\[
  \epsilon(\ell, d \ell, t) = 1 - \frac{q_{1}(\ell, t)}{q_{2}(\ell, d \ell, t)} \; .
\]

\subsection*{Susceptible-infected-recovered model}
For the SIR model, the calculation is a bit more cumbersome than for the SI model. 

Suppose node $s$ is initially in the infected state, and suppose that
two independent paths of length $\ell$ and $\ell + d \ell$ connect node $i$
to node $s$.
The probability $q_{2}(\ell, \ell + d \ell, t)$ that node $i$ becomes infected at time $t$
is given by the probability that the infection spreads along at \change{least} one
of these paths. We remark that we know the analytical form
of the probability $q_{1}(\ell, t)$ that the infection spreads along a single
path of length $\ell$ in $t$ time steps or less, see main text.
However, this expression can be used to combine the contribution
of the two independent paths only provided that the
paths are dynamically independent. The latter condition
is satisfied only when the infection performs
at least one step towards the target along at least one of the paths.

\begin{table*}[!htb]
    \begin{tabular}{|c|c|c|c|c|}
      \hline
    $\sigma_v$ & $\sigma_v$ & $\sigma_s$ & $\textrm{Prob.}(\mathbf{\sigma})$ & $q_{2}(\ell, \ell + d \ell, t |
                                    \mathbf{\sigma})$
      \\
    \hline
      \hline
      $S$ & $S$ & $I$ & $(1-\beta)^2 (1- \gamma)$ & $q_{2}(\ell, \ell + d \ell, t-1)$
      \\
      \hline
      $S$ & $S$ & $R$ & $(1-\beta)^2 \gamma$ & $0$
      \\
      \hline
      $S$ & $I$ & $I$ & $\beta (1-\beta) (1-\gamma)$ & $1 - [1-q_{1}(\ell,
                                                         t-1)]
                                                       [1-q_{1}(\ell + d \ell-1,
                                                         t-1)]$
      \\
      \hline
      $S$ & $I$ & $R$ & $\beta (1-\beta) \gamma$ & $q_{1} (\ell + d
                                                   \ell -1,
                                                         t-1)$
      \\
      \hline
      $I$ & $S$ & $I$ & $\beta (1-\beta) (1-\gamma)$ & $1 - [1-q_{1} (\ell-1,
                                                         t-1)]  [1-
                                                       q_{1} (\ell +
                                                       d \ell,
                                                         t-1)]$
      \\
      \hline
      $I$ & $S$ & $R$ & $\beta (1-\beta) \gamma$ & $q_{1} (\ell-1, t-1)$
      \\
      \hline
      $I$ & $I$ & $I$ & $\beta^2 (1-\gamma)$ & $1 - [1 - q_{1} (\ell-1,
                                                         t-1)]
                                               [1-q_{1} (\ell + d \ell-1,
                                                         t-1)]$
      \\
      \hline
      $I$ & $I$ & $R$ & $\beta^2 \gamma$ & $1- [1 - q_{1} (\ell-1,
                                                         t-1)] [1-
                                           q_{1} (\ell + d \ell -1,
                                                         t-1)]$
      \\
      \hline
  \end{tabular}
  \caption{Configurations $\mathbf{\sigma} = (\sigma_v, \sigma_w, \sigma_s, S , \dots , S)$ reachable after one dynamical step assuming
    that the configuration at preceding time is such that node $s$ is
    infected and all other nodes are susceptible. We provide the
    value of the probability $\textrm{Prob.}(\mathbf{\sigma})$ for
    each of these configurations to happen together with the conditional
    probability $q_{2}(\ell, \ell + d \ell, t |
                                    \mathbf{\sigma})$ that the infection will reach node $i$ along one of
    the two paths of length $\ell$ and $\ell + d \ell$, respectively. The latter probability is given by appropriate 
    combinations of the known probabilities $q_{1}$ for the single independent paths.}
  \label{tab:1}
\end{table*}

Indicate with $v$ the neighbor of node $s$ along the path of length
$\ell$ towards $i$, and with $w$ the neighbor of node $s$ along the path of length
$\ell + d \ell$ towards $i$. The initial configuration at time $t=0$
is such that $\sigma_s^{(0)} = I$ and $\sigma_{\forall j \neq s}^{(0)} = S$.  
At time $t=1$, the states of nodes may change as the results of
spreading and recovery events. The only nodes that can change their
states are $s$, $v$ and $w$.
For example, we can go to the configuration $\mathbf{\sigma}^{(1)} =
(I, I, S, \ldots)$, i.e.,  such that
$\sigma_v^{(1)} = I$, $\sigma_w^{(1)} =
S$ and $\sigma_s^{(1)} = I$, with probability
$\textrm{Prob.}[\mathbf{\sigma}^{(1)} = (\sigma_v^{(1)}  = I,
\sigma_w^{(1)} = S, \sigma_v^{(1)} = I, S, \ldots, S) ]
= \beta (1-\beta) (1-
\gamma)$. After this first step, the spreading of the infection will
happen independently along the two paths, thus we
can write
$q_2[\ell, \ell + d \ell, t | \mathbf{\sigma}^{(1)} = (\sigma_v^{(1)}  = I,
\sigma_w^{(1)} = S, \sigma_v^{(1)} = I, S, \ldots, S) ] = 1 - [1-q_1(\ell-1, t-1)]
[1-q_1(\ell + d \ell, t-1)]$. There are in total eight of such
configurations. They are listed in Table~\ref{tab:1}.
In general, we can
write that
\begin{equation}
q_2 (\ell, \ell + d \ell, t) = \sum_{\mathbf{\sigma}} \, q_2(\ell, \ell + d \ell, t | \mathbf{\sigma})  \, \textrm{Prob.}(\mathbf{\sigma}) \; ,
 \label{eq:SIR}
\end{equation}
where the sum runs over all eight configurations $\mathbf{\sigma}$
of Table~\ref{tab:1}. The expressions of the
probabilities appearing in Table~\ref{tab:1} are then used to solve
Eq.~(\ref{eq:SIR}) by iteration, starting from the
initial condition $q_2(\ell, \ell + d \ell, t =0 ) = 0$.

\section{Joint probability of infection from a single source}

\subsection*{Susceptible-infected model}

\begin{figure}[!hbt]
\centering
\includegraphics[width=0.4\textwidth]{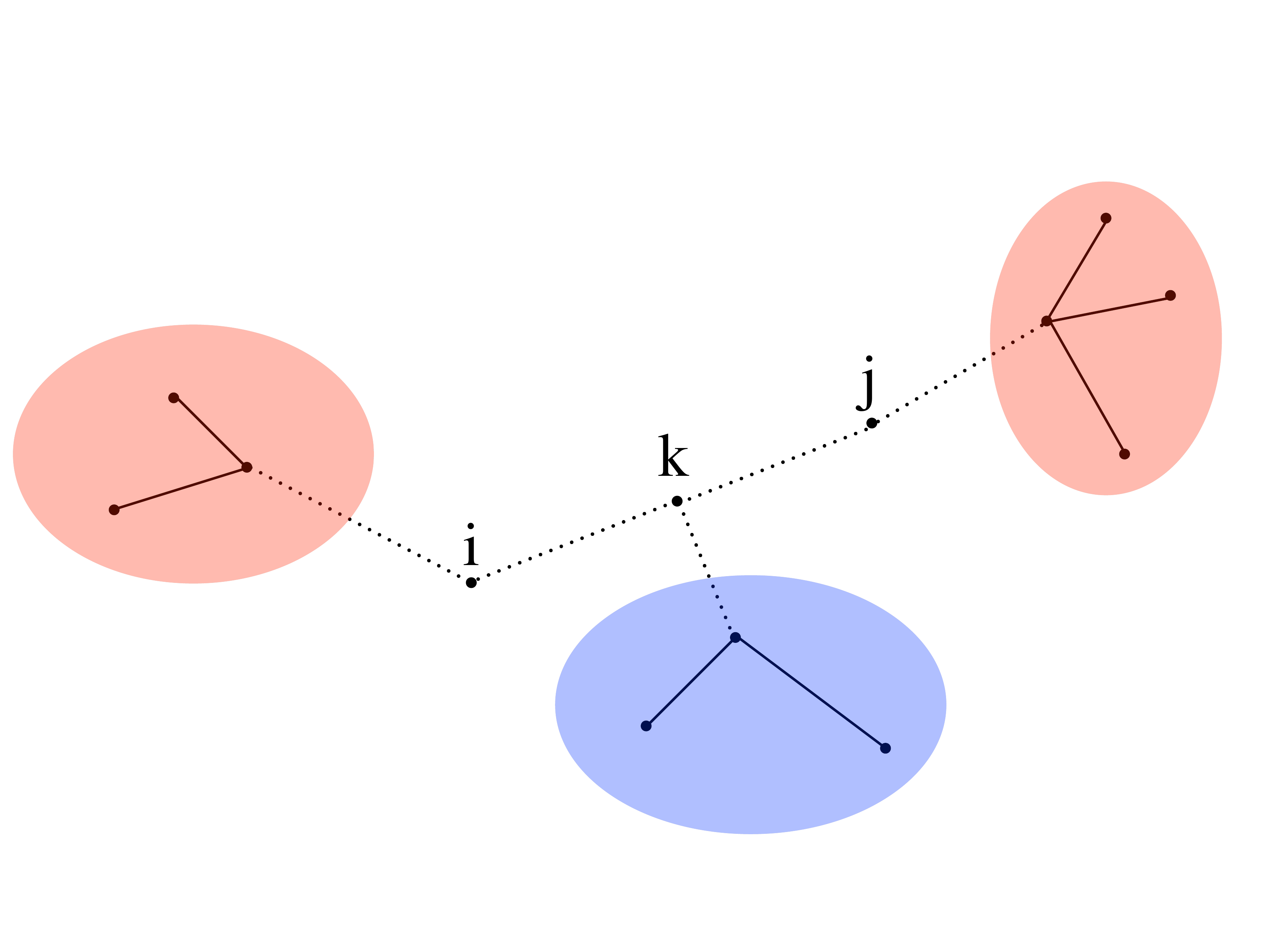}
\caption{Schematic illustration for the computation of the joint probability. The shaded areas highlight different parts of the network where the source node can be located, relatively to the positions of the target nodes $i$ and $j$. Red areas denote regions where one of the two paths of spreading is dependent on the other. The blue shaded area indicate locations of the source node leading to path of spreading that are partially independent.}
\label{fig:tree cases}
\end{figure}

Here, we illustrate how to compute the joint probability $Q^{(t)}_{s\to i,j}$ that nodes $i$ and $j$ are infected at time $t$ or earlier given that the source of spreading is node $s$. The computation still take\change{s} advantage of Eqs.~(\ref{eq:Q_sing}) and~(\ref{eq:P_sing}), by properly accounting for the position of the source node
$s$ relatively to the positions of the target nodes $i$ and $j$ (see Figure~\ref{fig:tree cases}). 

If node $j$ is seating in between nodes $s$ and $j$, then the infection can reach node $i$ only passing first through node $j$. Thus, we can safely write that $Q^{(t)}_{s \to i,j} = Q^{(t)}_{s \to i}$. The same exact argument leads us to write $Q^{(t)}_{s \to i,j} = Q^{(t)}_{s \to j}$ if node $i$ is seating in between nodes $j$ and $s$.

A less straightforward computation is required when the source node $s$ is connected to nodes $i$ and $j$ with partially independent paths. Part of the spreading path can be in common among the two trajectories, say up to node $k$ as indicated in Figure~\ref{fig:tree cases}. However after this node, the two paths are dynamically independent one on the other and the two contributions are computed separately. Specifically, we can write

\begin{equation}
Q^{(t)}_{s\to i,j} = \sum_{r=0}^{t-\max(\ell_{ki},\ell_{kj})} P^{(r)}_{s \to k} \, Q^{(t-r)}_{k\to i} \, Q^{(t-r)}_{k\to j} \; ,
\label{eq:joint_SI}
\end{equation}
where $P^{(r)}_{s \to k}$ is the usual probability that the infection reached node $k$ in exactly $r$ stages of the dynamics. The sum on the r.h.s. of Eq.~(\ref{eq:joint_SI}) runs over all possible values of $r$ compatible with the quantity that we want to estimate.

\subsection*{Susceptible-infected-recovered model}

In the SIR model we can compute $Q^{(t)}_{s\to i,j}$ using the very same method for SI with the only caveat to take into account Eq.~(\ref{eq:SIR}) and  Table~\ref{tab:1} whenever the source is between $i$ and $j$ or the two shortest paths become independent.

\bibliographystyle{apsrev}
\bibliography{bib}

\end{document}